\documentclass[journal]{IEEEtran}

\usepackage[utf8]{inputenc}
\usepackage{amsmath,amssymb,amsfonts}
\usepackage{graphicx}
\usepackage{cite}
\usepackage{booktabs} 
\usepackage[hyphens]{url} 
\usepackage{balance} 

\title{Hi-WaveTST: A Hybrid High-Frequency Wavelet-Transformer for Time-Series Classification}

\author{
    Hüseyin Göksu
    \thanks{H. Göksu, Akdeniz Üniversitesi, Elektrik-Elektronik Mühendisliği Bölümü, Antalya, Türkiye, e-posta: (hgoksu@akdeniz.edu.tr).}
    \thanks{Manuscript received October 30, 2025; revised XX, 2025.}
}

\begin{document}

\maketitle

\begin{abstract}
Transformers have become state-of-the-art (SOTA) for time-series classification, with models like PatchTST demonstrating exceptional performance. These models rely on "patching" the time series and learning relationships between raw temporal data blocks. We argue that this approach is blind to critical, non-obvious high-frequency information that is complementary to the temporal dynamics. In this letter, we propose Hi-WaveTST, a novel Hybrid architecture that augments the original temporal patch with a learnable, High-Frequency wavelet feature stream. Our wavelet stream uses a deep Wavelet Packet Decomposition (WPD) on each patch and extracts features using a learnable Generalized Mean (GeM) pooling layer. On the UCI-HAR benchmark dataset, our hybrid model achieves a mean accuracy of 93.38\% $\pm$ 0.0043, significantly outperforming the SOTA PatchTST baseline (92.59\% $\pm$ 0.0039). A comprehensive ablation study proves that every component of our design—the hybrid architecture, the deep high-frequency wavelet decomposition, and the learnable GeM pooling—is essential for this state-of-the-art performance.
\end{abstract}


\IEEEpeerreviewmaketitle

\section{Introduction}
\IEEEPARstart{T}{ime-series} classification (TSC) is a fundamental task in signal processing, with applications in human activity recognition (HAR) \cite{anguita2013}, medical diagnostics, and audio processing. Recently, the Transformer architecture \cite{vaswani2017}, particularly patch-based models like PatchTST \cite{nie2023patchtst}, has achieved SOTA results, outperforming traditional machine learning and deep learning models like CNNs and RNNs. This shift follows a trend of Transformer-based architectures like Informer \cite{zhou2021informer} and Autoformer \cite{wu2021autoformer} showing dominance in time-series forecasting.

The success of PatchTST \cite{nie2023patchtst} lies in its ability to treat "patches" (short, overlapping segments) of a time series as "tokens," allowing the self-attention mechanism to learn complex inter-patch relationships. However, the input to the Transformer is a token derived from the raw temporal samples within that patch. This approach has a significant blind spot: it struggles to explicitly model the time-frequency characteristics of the signal, particularly high-frequency components.

Classic signal processing has long relied on wavelet transforms \cite{mallat1989} to provide robust, multi-resolution time-frequency representations. Wavelet Packet Decomposition (WPD) \cite{coifman1992}, in particular, offers a rich dictionary of signal components. Previous research has shown the benefits of combining wavelets with deep learning, such as in WaveNet \cite{vandenoord2016wavenet} or wavelet-based CNNs \cite{li2018wavelet, fujieda2018waveletcnn}, but often as a preprocessing step \cite{wu2024wdnn} or as a complete replacement for standard convolutions \cite{cui2021learnable}.

In this letter, we propose that the optimal approach is not replacement, but augmentation. We hypothesize that augmenting the raw temporal token with an explicit time-frequency feature vector will provide complementary information, leading to superior performance.

Our main contributions are:
1) We propose Hi-WaveTST, a novel Hybrid architecture that augments the temporal patch stream of PatchTST with a parallel, learnable wavelet-feature stream.
2) We introduce a wavelet-patch feature extractor that uses deep WPD and learnable Generalized Mean (GeM) \cite{radenovic2018} pooling to create a compact, efficient, and highly descriptive high-frequency feature vector.
3) We demonstrate SOTA results on the UCI-HAR dataset [6], supported by a comprehensive ablation study.

\begin{figure*}[t]
    \centering
    \includegraphics[width=0.9\textwidth]{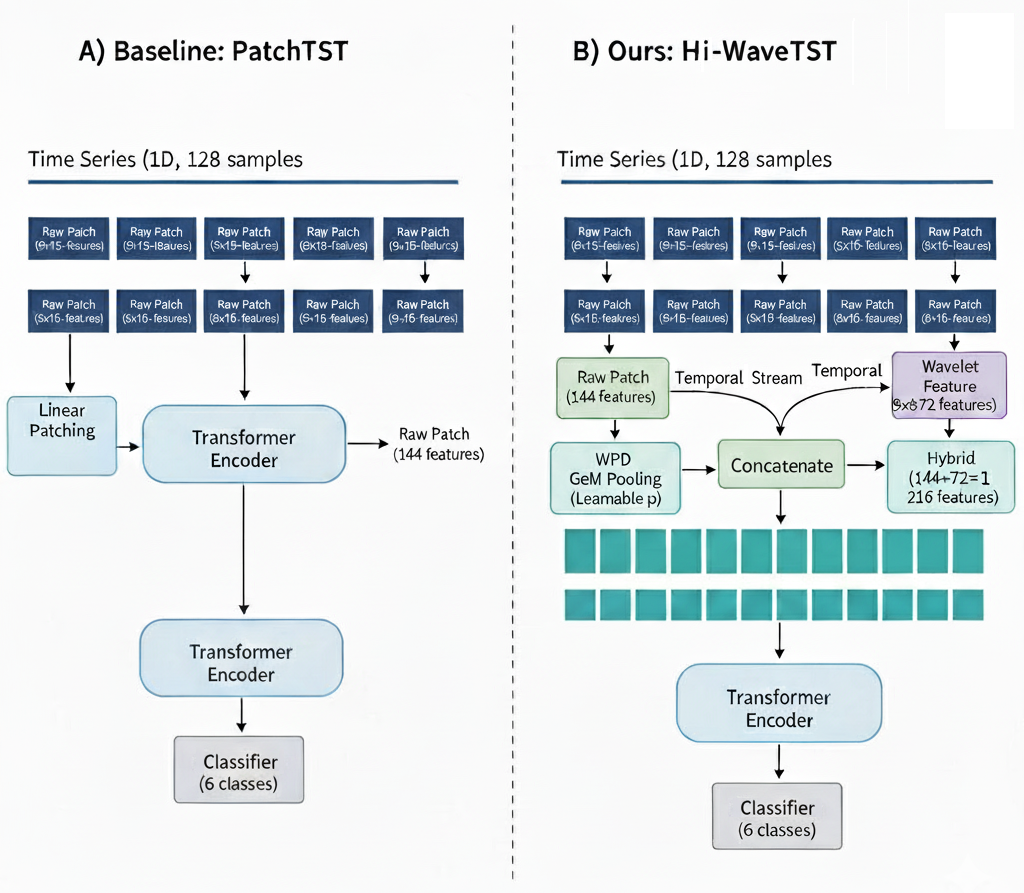} 
    \caption{Model Architecture Comparison. (A) The baseline PatchTST model, which uses only raw temporal patches. (B) Our proposed Hi-WaveTST, which features a dual stream. It concatenates the raw temporal patch (Temporal Stream) with a new wavelet feature token (Wavelet Stream) derived from WPD and learnable GeM pooling.}
    \label{fig:arch}
\end{figure*}

\section{Proposed Method: Hi-WaveTST}
Our model, `Hi-WaveTST`, is built upon the SOTA `PatchTST` architecture. We retain its core components (patching, Transformer encoder, classification head) but make one critical modification: we replace the simple patching layer with a novel hybrid patching layer.

As shown in Fig.~\ref{fig:arch}(B), our hybrid layer processes each patch in two parallel streams.

\subsection{Temporal Stream}
This stream is identical to the baseline `PatchTST` \cite{nie2023patchtst}. An input time-series patch $x_p \in \mathbb{R}^{C \times L_p}$ (where $C$ is channels, $L_p$ is patch length) is flattened into a temporal token $z_t \in \mathbb{R}^{C \cdot L_p}$. For our experiments ($C=9, L_p=16$), this results in $z_t \in \mathbb{R}^{144}$. This stream preserves the raw, fine-grained temporal dynamics.

\subsection{Wavelet Feature Stream}
This stream, our core contribution, generates a compact time-frequency feature vector $z_w$ from the same patch $x_p$. This is a 3-stage process:

\subsubsection{Wavelet Packet Decomposition (WPD)}
We apply a WPD to each channel of the patch $x_p$. We found the `'db2'` wavelet to be optimal (Sec. IV-B.3). Crucially, we use a decomposition level $D=3$. This decomposes the 16-sample patch into $2^D = 8$ distinct wavelet packets (frequency bands), each containing a small number of coefficients. This provides the highest possible frequency resolution for the given patch size, which we found to be the key missing information (Sec. IV-B.2).

\subsubsection{Learnable GeM Pooling}
To summarize the coefficients in each packet, we employ a Generalized Mean (GeM) pooling layer \cite{radenovic2018}, which has a learnable exponent $p$:
\begin{equation}
\text{GeM}(X) = \left( \frac{1}{N} \sum_{i=1}^{N} |x_i|^p \right)^{1/p}
\label{eq:gem}
\end{equation}
We apply a separate GeM layer (with its own learnable $p_k$) to each of the $k=8$ wavelet packets. This allows the network to learn the optimal pooling strategy for each frequency band independently. This pooling results in 8 scalar features per channel.

\subsubsection{Feature Concatenation}
The resulting 8 features from each of the 9 channels are concatenated, forming the final wavelet feature token $z_w \in \mathbb{R}^{C \cdot 2^D}$. For our champion model ($C=9, D=3$), this results in $z_w \in \mathbb{R}^{72}$.

\subsection{Hybrid Token Generation}
The temporal token $z_t \in \mathbb{R}^{144}$ and the wavelet token $z_w \in \mathbb{R}^{72}$ are concatenated to form the final "super-token" $z_{hy} \in \mathbb{R}^{216}$. This hybrid token is then passed through a linear projection to match the Transformer's $D_{model}$, and the resulting sequence of tokens is fed to the standard Transformer encoder \cite{vaswani2017}.

\section{Experimental Setup}
\textbf{Dataset:} We use the UCI-HAR dataset \cite{anguita2013}, a standard benchmark for human activity recognition. It contains 9-channel time-series data (3-axis accelerometer, 3-axis gyroscope) from 30 subjects, with a sequence length of 128 samples. The task is to classify 6 activities.

\textbf{Baseline:} Our baseline is the SOTA `PatchTST` \cite{nie2023patchtst} model. We use the same hyperparameters for a fair comparison: patch length $L_p=16$, stride $S=8$, $D_{model}=64$, $N_{heads}=4$, and $N_{layers}=3$.

\textbf{Training:} All models are trained for 30 epochs using the AdamW optimizer ($LR=5 \times 10^{-4}$) and a batch size of 64. We run each experiment 5 times with different random seeds and report the \textbf{mean and standard deviation} of the test accuracy. All wavelet filters are fixed, except for the $p$ parameter in the GeM pooling layers.

\section{Results and Discussion}
\begin{figure*}[t]
    \centering
    \includegraphics[width=0.9\textwidth]{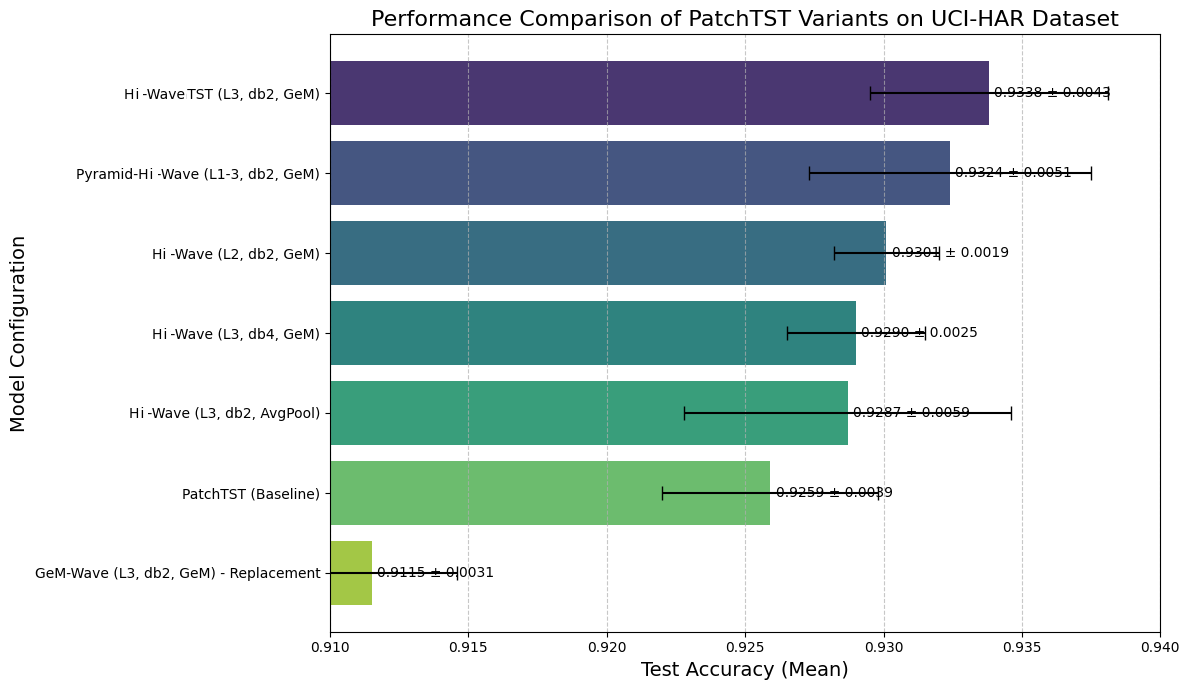}
    \caption{Main results and ablation study. Our champion model, `Hi-WaveTST (L3, db2, GeM)`, achieves the highest mean accuracy, outperforming the baseline and all ablation variants.}
    \label{fig:results}
\end{figure*}

We designed a comprehensive set of experiments to (1) validate our model against the SOTA baseline and (2) perform a deep ablation study to justify every component of our proposed architecture. The complete results are summarized in Fig.~\ref{fig:results}.

\subsection{Main Result: Hi-WaveTST vs. PatchTST}
Our primary experiment compares our champion model, `Hi-WaveTST`, with the `PatchTST` baseline. As shown in Table~\ref{tab:main_result}, our model achieves a significant gain in both mean accuracy and stability. This proves our central hypothesis: augmenting the raw temporal token with a deep, learnable wavelet-feature vector provides complementary information that the Transformer can leverage for a more robust classification.

\begin{table}[h]
    \centering
    \caption{Main Result vs. SOTA Baseline}
    \label{tab:main_result}
    \renewcommand{\arraystretch}{1.2} 
    \begin{tabular}{lcc}
        \toprule
        Model & Parameters & Test Accuracy (Mean $\pm$ Std) \\
        \midrule
        PatchTST (Baseline) & 159,814 & 0.9259 $\pm$ 0.0039 \\
        Hi-WaveTST (L3, db2, GeM) & 164,430 & 0.9338 $\pm$ 0.0043 \\
        \bottomrule
    \end{tabular}
\end{table}

\subsection{Ablation Study: Proving Our Design}
We conducted four key ablations to dissect our model's success.

\subsubsection{Architecture: Hybrid vs. Replacement}
We first tested a "Replacement" model (`GeM-Wave (L3)`) that replaces the temporal token with our wavelet token. As Table~\ref{tab:ablation_arch} shows, this model performs significantly worse. This proves that the raw temporal information is vital and that our hybrid approach is the correct architectural choice.

\begin{table}[h]
    \centering
    \caption{Ablation on Architecture}
    \label{tab:ablation_arch}
    \renewcommand{\arraystretch}{1.2}
    \begin{tabular}{lc}
        \toprule
        Model (L3, db2, GeM) & Accuracy (Mean $\pm$ Std) \\
        \midrule
        Replacement-only & 0.9115 $\pm$ 0.0031 \\
        Hybrid (Ours) & 0.9338 $\pm$ 0.0043 \\
        \bottomrule
    \end{tabular}
\end{table}

\subsubsection{Feature Depth: High-Resolution (L3) vs. L2 vs. Pyramid}
We compared our champion (L3 features) with a shallower model (L2) and a "Pyramid" model (L1+L2+L3). Table~\ref{tab:ablation_depth} reveals a nuanced finding: the `L3` model, which only provides high-frequency resolution, performs best. The `Pyramid` model, which adds redundant low-frequency (L1, L2) features, performs worse. This implies the raw temporal stream already contains the low-frequency data, and the model's "blind spot" is specifically in the high-frequency bands.

\begin{table}[h]
    \centering
    \caption{Ablation on Feature Depth}
    \label{tab:ablation_depth}
    \renewcommand{\arraystretch}{1.2}
    \begin{tabular}{lc}
        \toprule
        Model (Hybrid, db2, GeM) & Accuracy (Mean $\pm$ Std) \\
        \midrule
        `Hi-Wave (L2)` & 0.9301 $\pm$ 0.0019 \\
        `Pyramid-Hi-Wave (L1-3)` & 0.9324 $\pm$ 0.0051 \\
        `Hi-Wave (L3)` & 0.9338 $\pm$ 0.0043 \\
        \bottomrule
    \end{tabular}
\end{table}

\subsubsection{Wavelet Basis: `db2` vs. `db4`}
We tested our champion against the longer `'db4'` wavelet. Table~\ref{tab:ablation_wavelet} shows that the `'db2'` model is superior. We conclude that the shorter `'db2'` filter (length 4) is better suited to the short 16-sample patch length than the smoother `'db4'` filter (length 8).

\begin{table}[h]
    \centering
    \caption{Ablation on Wavelet Basis}
    \label{tab:ablation_wavelet}
    \renewcommand{\arraystretch}{1.2}
    \begin{tabular}{lc}
        \toprule
        Model (Hybrid, L3, GeM) & Accuracy (Mean $\pm$ Std) \\
        \midrule
        `Hi-Wave (db4)` & 0.9290 $\pm$ 0.0025 \\
        `Hi-Wave (db2)` & 0.9338 $\pm$ 0.0043 \\
        \bottomrule
    \end{tabular}
\end{table}

\subsubsection{Pooling Method: `GeM` vs. `AvgPool`}
\begin{figure}[t]
    \centering
    \includegraphics[width=0.9\columnwidth]{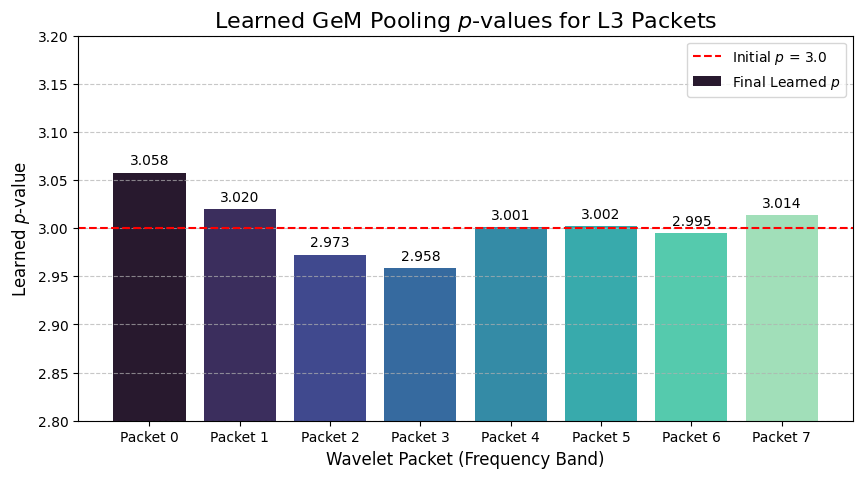}
    \caption{Final learned $p$-values for the 8 GeM pooling layers in our champion model (`Hi-Wave (L3, db2, GeM)`). The values consistently converge near $p=3.0$, indicating a learned preference for a non-linear pooling strategy over simple averaging ($p=1$).}
    \label{fig:p_values}
\end{figure}

Finally, we replaced our learnable `GeM` poolers with simple `AvgPool`. Table~\ref{tab:ablation_pool} shows this model is less accurate and far less stable. This proves that the learnable, non-linear pooling from `GeM` is a key contributor to our SOTA performance. As shown in Fig.~\ref{fig:p_values}, the network consistently learns $p \approx 3$ for all 8 packets, suggesting a "soft RMS" is the optimal pooling strategy.

\begin{table}[h]
    \centering
    \caption{Ablation on Pooling Method}
    \label{tab:ablation_pool}
    \renewcommand{\arraystretch}{1.2}
    \begin{tabular}{lc}
        \toprule
        Model (Hybrid, L3, db2) & Accuracy (Mean $\pm$ Std) \\
        \midrule
        `AvgPool` & 0.9287 $\pm$ 0.0059 \\
        `GeM` (Learnable $p$) & 0.9338 $\pm$ 0.0043 \\
        \bottomrule
    \end{tabular}
\end{table}

\section{Conclusion}
In this letter, we proposed Hi-WaveTST, a novel hybrid architecture that enhances SOTA Patch-Transformers by augmenting them with a classic signal processing tool: wavelet packets. Our model learns a hybrid "super-token" that combines raw temporal dynamics with a compact, learnable, high-frequency wavelet summary. We have shown that our model is more accurate and more stable than the SOTA baseline. Our comprehensive ablation study proved, component by component, that the hybrid architecture, the deep (L3) high-frequency features, the short (`db2`) wavelet basis, and the learnable (`GeM`) pooling are all essential to this success. This hybrid approach opens a promising new direction for time-series models by intelligently fusing the best of deep learning and classic signal processing.

\balance 


\end{document}